\newcommand{\p}{\partial}
\newcommand{\pslash}{p\kern-1ex /}
\newcommand{\lslash}{l\kern-1ex /}
\newcommand{\kslash}{k\kern-1ex /}
\newcommand{\dslash}{\p\kern-1.2ex /}
\newcommand{\Dslash}{{\cal D}\kern-1.5ex /}

\newcommand{\bea}{\begin{eqnarray}}
\newcommand{\eea}{\end{eqnarray}}

\newcommand{\BAN}{\begin{eqnarray*}}
\newcommand{\EAN}{\end{eqnarray*}}
\documentclass{PoS}
\usepackage{graphicx}
\usepackage{booktabs}

\title{One-Flavor Algorithms for Simulation of Lattice QCD with Domain-Wall Fermion: EOFA versus RHMC}

\ShortTitle{{\rm EOFA versus RHMC}}

\author{\speaker{Yu-Chih Chen}$^{1}$, 
        Ting-Wai Chiu$^{1,2}$ (TWQCD Collaboration)\\
$^1$ Physics Department, National Taiwan University, Taipei 10617, Taiwan\\
$^2$ Center for Quantum Science and Engineering, National Taiwan University,\\
     \hspace{2.5mm}Taipei 10617, Taiwan\\
     }

\abstract{
We compare the performances of the exact one-flavor algorithm (EOFA) and the rational hybrid Monte Carlo algorithm (RHMC),     
for dynamical simulations of lattice QCD with domain-wall fermion. 
}

\FullConference{2013 International Workshop on Computational Science and Engineering,\\
		14-17 October 2013,\\
		National Taiwan University, Taipei, Taiwan}

\begin{document}

\section{Introduction}

Recently, an exact pseudofermion action for hybrid Monte Carlo simulation (HMC) of lattice QCD 
with one-flavor of domain-wall fermion (DWF) has been derived, 
with the effective 4-dimensional Dirac operator equal to the optimal rational approximation 
of the overalp Dirac operator with kernel $ H = c H_w (1 + d \gamma_5 H_w)^{-1} $, where $ c $ and $ d $ are constants, 
and $ H_w $ is the standard Wilson-Dirac operator plus a negative parameter $ -m_0 $ ($ 0 < m_0 < 2 $) \cite{Chen:2014hyy}.
Since the action is exact without taking square root, it does not require a large memory space to compute the fermion force,
unlike the widely used rational hybrid Monte Carlo algorithm (RHMC) \cite{Clark:2006fx}.
In the following, we refer the HMC with the exact one-flavor pseudofermion action as the exact one-flavor algorithm (EOFA).
Obviously, the memory-saving feature of EOFA is crucial for large-scale simulations of lattice QCD on any platforms. 
This is especially true for GPUs, since each GPU has enormous floating-point computing power but limited device memory.
For example, using EOFA, two GPUs (each of 6 GB device momory, e.g., Nvidia GTX-TITAN)
working together is capable to simulate lattice QCD with $ (u,d,s,c) $ DWF quarks on the
$ 32^3 \times 64 \times 16 $ lattice, while this is infeasible for RHMC. 
About the computational efficiency of EOFA, our studies in Ref. \cite{Chen:2014hyy} suggest that EOFA is compatible with RHMC. 
However, in Ref. \cite{Chen:2014hyy}, a salient feature of EOFA has not been exploited. Namely, 
$ \phi_1 $ and $ \phi_2 $ [see Eq. (23) of Ref. \cite{Chen:2014hyy}] can be updated at two different time scales,  
since the fermion force of $ \phi_1 $ is much smaller than that of $ \phi_2 $. 
Now, applying the multiple-time scale method to $ \phi_1 $ and $ \phi_2 $, we find that EOFA is more efficient than
RHMC for all variants of DWF. 
In this paper, we demonstrate that this is the case for the conventional DWF with
kernel $ H = 2 c H_w ( 2 + \gamma_5 H_w )^{-1} $ (i.e, $ d = 1/2 $, and $\omega = 1 $), 
and the optimal DWF \cite{Chiu:2002ir} with kernel $ H = H_w $ (i.e., $ c = 1 $, $ d = 0 $, 
and the reflection-symmetric $ \omega $),   
and the tests are performed for $N_f=1$ and $ N_f=(2+1)$ QCD respectively. 

Since the details of EOFA have been given in Ref. \cite{Chen:2014hyy}, we do not repeat them here.
In the following, we outline how we implement RHMC in our tests.
The pseudofermion action for RHMC of $N_f=1$ DWF can be written as  
\bea
\label{eq:Spf_RHMC}
S_{pf}^{N_f=1} = \phi^\dagger (C_1^\dagger C_1)^{1/4} (C C^\dagger)^{-1/2} (C_1^\dagger C_1)^{1/4} \phi, 
\eea
where $ C $ is defined as \cite{Chiu:2013aaa},  
\BAN
\label{eq:Cm}
C(m) &\equiv& I-M_5(m)D_w^{oe}M_5(m)D_w^{eo}, \hspace{2mm} C_1 \equiv C(1) \\
\label{eq:DwEO}
(D_w^{eo(oe)})_{x,y}
&\equiv& \frac{1}{2}[(\gamma_{\mu}-1)U_{\mu}(x)\delta_{y,x+\mu}-(\gamma_{\mu}+1)U_{\mu}^{\dag}(y)\delta_{y,x-\mu}], \\
\label{eq:M5}
M_5(m) &\equiv& [4-m_0+P_+M_+(m)+P_-M_-(m)]^{-1}, 
\EAN
and the number of poles in the optimal rational approximation of $ (C C^\dagger)^{-1/2} $ and $ (C_1^\dagger C_1)^{1/4} $ is $N_p$. 
The pseudofermion field $\phi$ is generated from the Gaussian noise field $\eta$ as follows.  
\BAN
\phi = \frac{1}{[C(1)C^{\dag}(1)]^{1/4}}[C(m)C^{\dag}(m)]^{1/4}\eta.
\EAN
At this point, we note that one can use the following DWF action to reduce the memory consumption in RHMC. 
\BAN
S_{pf}^{N_f=1} = \Phi^\dagger (C_1^\dagger C_1)^{1/2} \Phi + \phi^\dagger (C^\dagger C)^{-1/2} \phi,  
\EAN
where $ \Phi $ and $ \phi $ are independent pseudofermion fields.
Then the fermion force due to $N_p$ poles 
can be computed in $ n $ subsets, each with multiple-shift CG. 
Thus the memory consumption can be reduced by a factor of $ n $. 
However, it takes more time to compute these $n$ subsets than just one set with $ N_p $ poles. 
To save time, one may apply the multiple-time scale method to these $n$ subsets.
Nevertheless, one cannot apply the mass preconditioning for this action, which may be a drawback of this approach. 
In the following, we use (\ref{eq:Spf_RHMC}) for RHMC in all tests.

\section{Memory Requirements for EOFA and RHMC}

Defining $M_{S} = L^3 T \times [\textmd{8 bytes (for double precision real number)}]$, then 
the link variables (with each $SU(3)$ matrix in the format of 2-column storage) take 
$ M_U = 48 M_S$,   
the conjugate momenta 
$ M_P = 32 M_S $, 
and a 5D vector (on the 5-dimensional lattice)  
$ M_V = 24 N_s M_S $, 
where $ N_s $ is the extent in the fifth dimension.

For EOFA, it takes $2 M_U$ to store the old and new gauge configurations,    
$M_P$ for the conjugate momenta, $ 2 \times 24 M_S$ for $\phi_1$ and $\phi_2$ (pseudofermion fields) of each fermion, and  
$M_{P}$ for the fermion force. 
To compute the fermion force by conjugate gradient, it needs $3.5 \times M_V $ for the working space.  
Thus the memory requirement for EOFA (with one heavy mass preconditioner) amounts to
\bea
\label{eq:mem_EOFA}
M_{\rm EOFA} = 2M_{U}+2M_{P}+2 \times 48 M_S + 3.5 \times M_V = 8( 32 + 10.5 N_s) M_{S}.
\eea

For RHMC, it takes $ 2 M_U$ (for old and new gauge configurations), $2 M_P$ (for conjugate momenta and fermion force), and   
$2 \times 12 N_s M_S $ for the pseudofermion fields (the light fermion and the heavy mass preconditioner) after taking into account 
of even-odd preconditioning.
To compute the fermion force, it needs $(2+2N_p)$ 5D vectors for mult-shift CG and working space, 
where $ N_p $ is the number of poles used in the rational approximation.
Thus the memory requirement for RHMC is
\bea
\label{eq:mem_RHMC}
M_{\rm RHMC} = 2M_{U}+2M_{P}+(3+2N_p) (24 N_s M_S) = 8[20 + 3(3+2 N_p) N_s ] M_S. 
\eea

From (\ref{eq:mem_EOFA}) and (\ref{eq:mem_RHMC}), the ratio is
\BAN
\label{eq:memory_ratio}
\frac{M_{\rm RHMC}}{M_{\rm EOFA}} = \frac{20 + 3 (3 + 2 N_p) N_s }{32 + 10.5 N_s},   
\EAN
independent of the size of the 4D lattice. For example, for $ N_p = 12 $ and $ N_s = 16 $, the ratio is 6.58. 
In other words, for HMC of one-flavor QCD with DWF on the $ 32^3 \times 64 \times 16 $ lattice, 
EOFA takes 12 GB, while RHMC with $N_p = 12 $ requires at least 79 GB.
Obviously, the memory-saving feature of EOFA has significant impacts to large-scale 
simulations on any platforms, especially for GPUs.

\section{Computational Efficiencies of EOFA and RHMC}

To compare the efficiencies of EOFA and RHMC, we perform the following tests: 
\begin{enumerate}
  \item 
         $N_f = 1 $ QCD on the $ 8^3 \times 16 \times 16 $ lattice 
  \begin{enumerate}
    \item  Conventional DWF with kernel $ H = 2 H_w (2 + \gamma_5 H_w)^{-1} $ (i.e., $ c = 1 $, $ d=1/2 $, and $ \omega = 1 $) 
           and $ m_0 = 1.8 $.  
    \item  Optimal DWF with kernel $ H = H_w $ (i.e., $ c = 1 $, $ d = 0 $, and the reflection-symmetric $ \omega $  
           with $ \lambda_{min}/\lambda_{max} = 0.05/6.2 $) and $ m_0 = 1.3 $.
  \end{enumerate}
  \item $ N_f = 1 $ and $ N_f = (2+1) $ QCD on the $ 16^3 \times 32 \times 16 $ lattice  
  \begin{enumerate}
    \item  Conventional DWF with kernel $ H = H_w (2 + \gamma_5 H_w)^{-1} $ (i.e., $ c = 1/2 $, $ d=1/2 $, and $ \omega = 1 $) 
           and $ m_0 = 1.8 $.  
  \end{enumerate}
\end{enumerate}

In all cases, the gauge action is the Wilson plaquette action at $ \beta = 6/g^2 = 5.95 $.  
In the molecular dynamics, we use the Omelyan integrator \cite{Omelyan:2001}, 
the multiple-time scale method \cite{Sexton:1992nu}, and the auxiliary heavy fermion field \cite{Hasenbusch:2001ne}. 
For $ N_f=1 $ QCD, the sea-quark mass is set to $ m_q a = 0.01 $, with the heavy mass preconditioner $ m_H a = 0.1 $ for 
conventional (optimal) DWF. 
For $ N_f=(2+1)$ QCD, the sea-quark masses are set to $ m_{u/d} a = 0.003 $   
with the heavy mass preconditioner $ m_H a = 0.03 $, and the values of $ m_s a = 0.01 $ and its mass preconditioner 
$ m_{H_s} a = 0.1 $. 
In RHMC, the number of poles in the optimal rational approximation of $ (C C^\dagger)^{-1/2} $ and $ (C_1^\dagger C_1)^{1/4} $ 
is fixed to $N_p=12$ for the lattice size $ 8^3 \times 16 \times 16 $, while $ N_p = 14 $ for $ 16^3 \times 32 \times 16 $.

\subsection{$L^3 \times T = 8^3 \times 16 $}

\begin{figure*}[tb]
\begin{center}
\begin{tabular}{@{}c@{}c@{}}
\includegraphics*[width=8cm,clip=true]{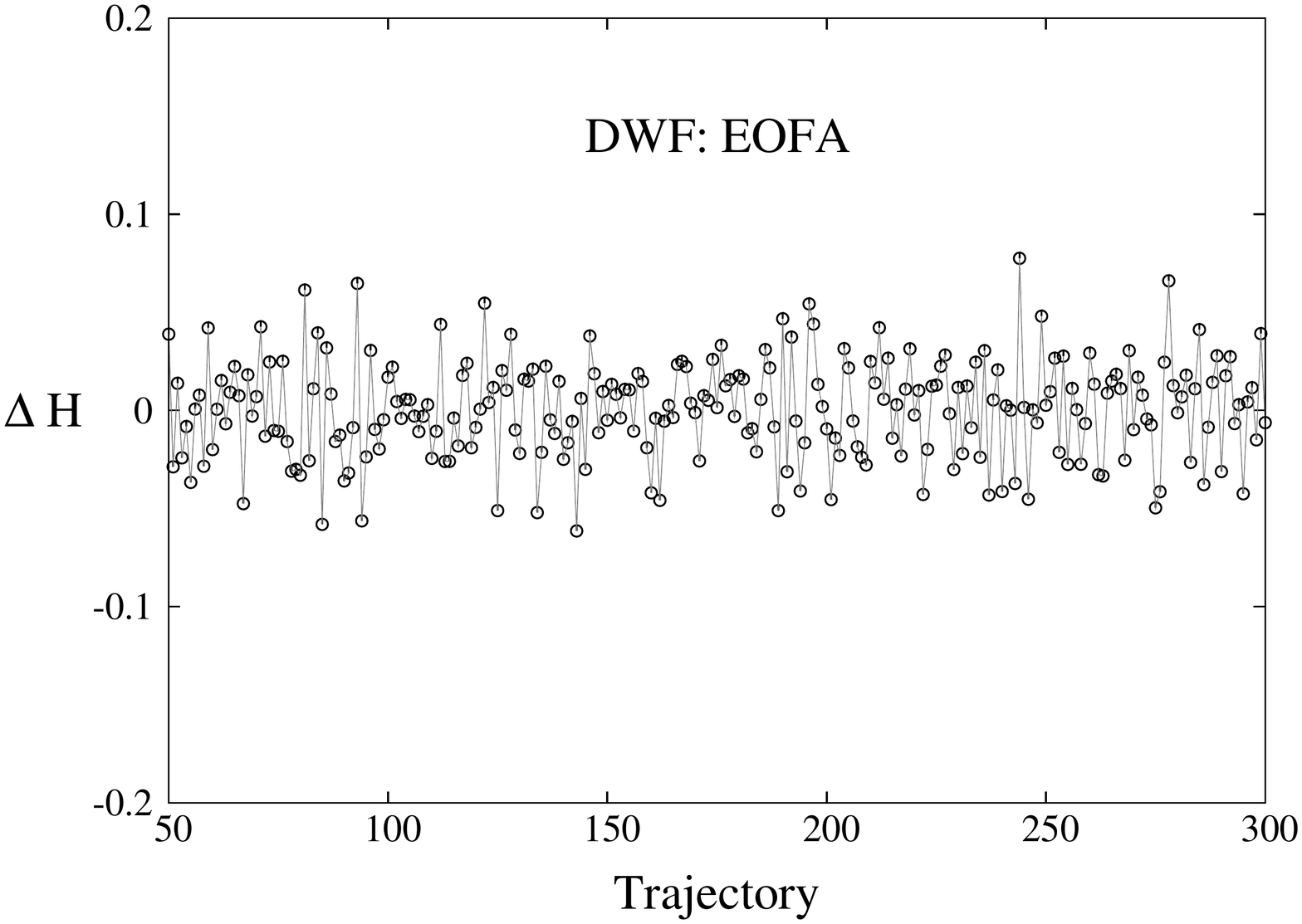}
&
\includegraphics*[width=8cm,clip=true]{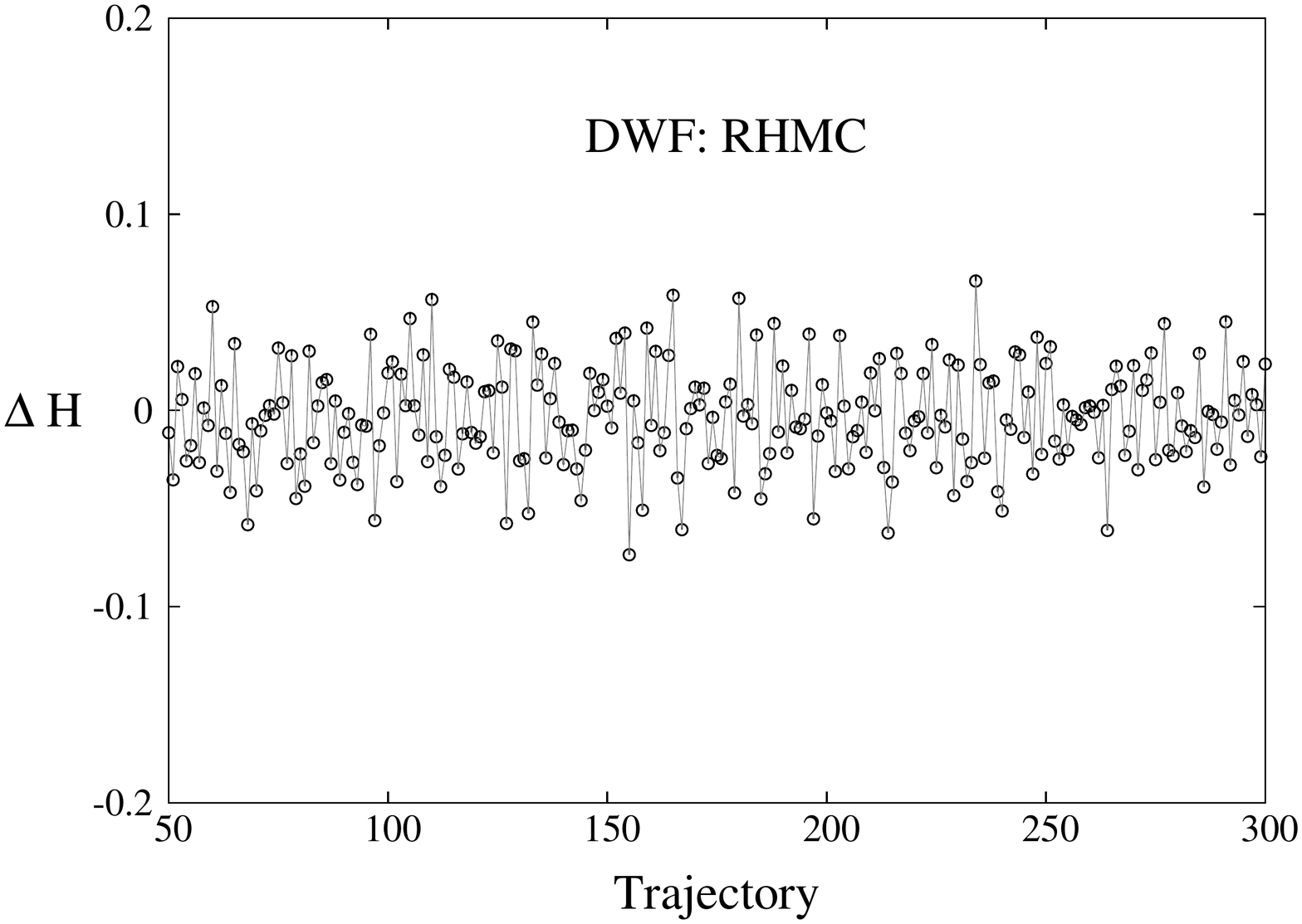}
\\ (a) & (b)
\end{tabular}
\caption{The change of Hamiltonian $ \Delta H $ versus the trajectory in the HMC of one-flavor QCD 
         with the conventional DWF, for (a) EOFA, and (b) RHMC respectively.
         The line connecting the data points is only for guiding the eyes.}
\label{fig:DeltaH_DWFs2}
\end{center}
\end{figure*}

\begin{figure*}[tb]
\begin{center}
\begin{tabular}{@{}c@{}c@{}}
\includegraphics*[width=8cm,clip=true]{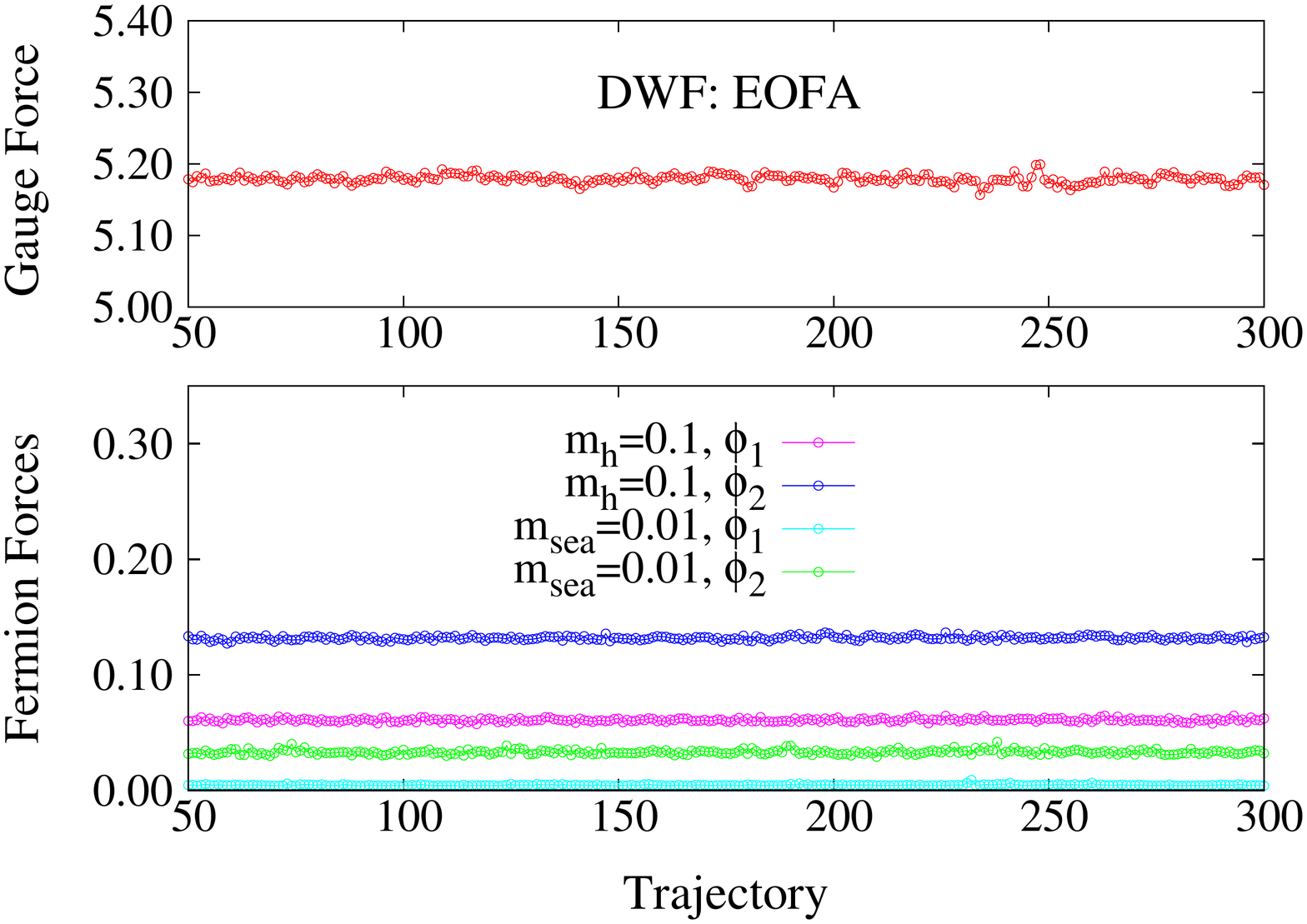}
&
\includegraphics*[width=8cm,clip=true]{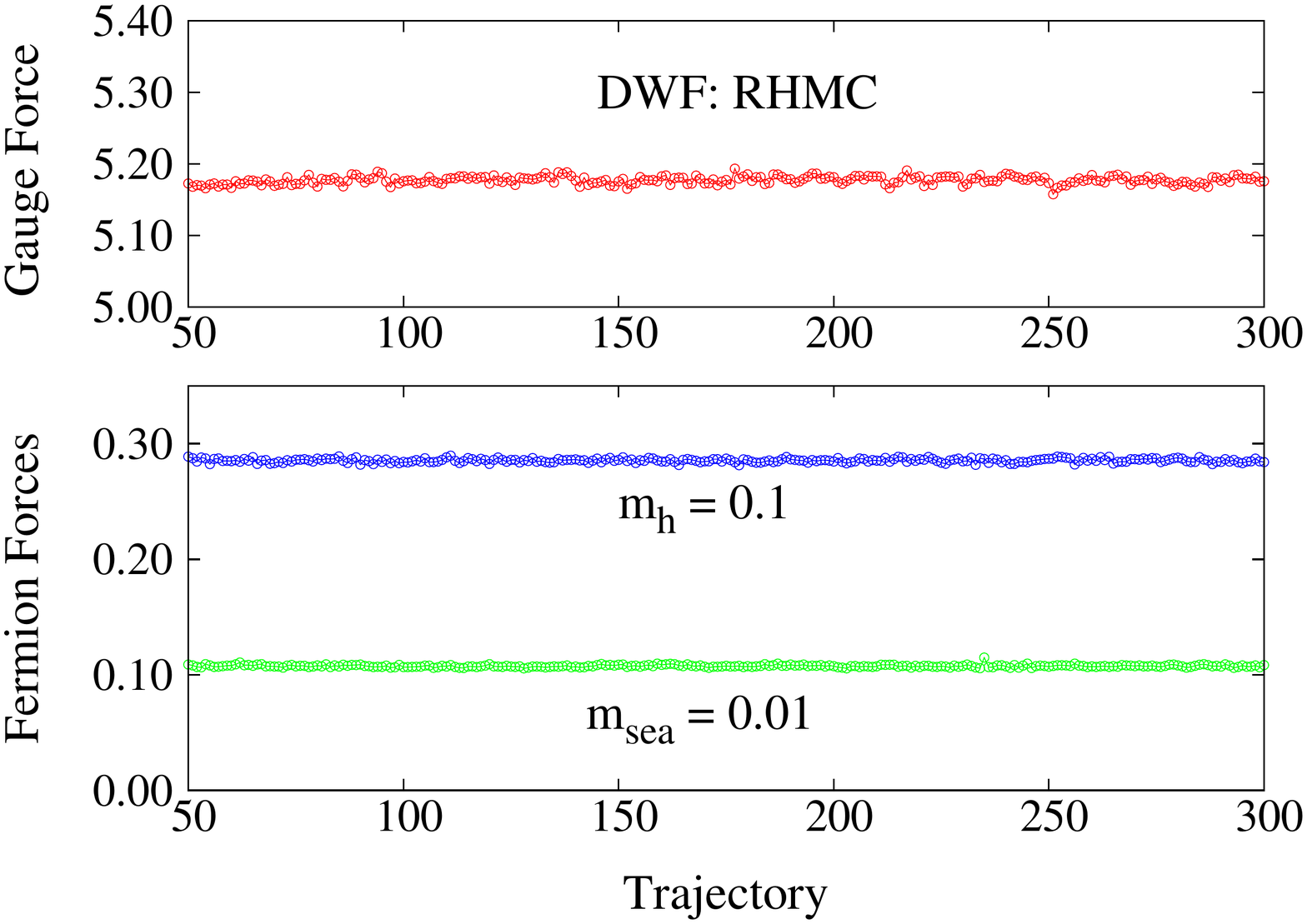}
\\ (a) & (b)
\end{tabular}
\caption{The maximum forces of the gauge field and the pseudofermion fields   
         versus the trajectory in the HMC of one-flavor QCD: (a) EOFA, (b) RHMC.}
\label{fig:HMC_forces_DWFs2}
\end{center}
\end{figure*}

First, we compare the HMC chracteristics of EOFA and RHMC.
In Fig. \ref{fig:DeltaH_DWFs2}, 
we plot the change of Hamiltonian $ \Delta H $ versus the trajectory number (after thermalization), for EOFA and RHMC respectively. 
In both cases, $ \Delta H $ is quite smooth without any spikes.
Moreover, the measured values of $ \left< \exp(-\Delta H) \right> $ are: 
\begin{center}
  \begin{tabular}{| l || c | r |}
    \hline
      & EOFA & RHMC \\ \hline
    Conventional DWF & 1.0003(16) & 1.0038(17) \\ \hline
    Optimal DWF & 0.9991(17) & 0.9994(18) \\ \hline
  \end{tabular}
\end{center}
They are all in good agreement with the condition $ \left< \exp(-\Delta H) \right> = 1 $
which follows from the area-preserving property of HMC.

In Fig. \ref{fig:HMC_forces_DWFs2}, 
we plot the maximum force (averaged over all links) in each trajectory, for the gauge field, the heavy fermion field,
and the light fermion field respectively. 
For both EOFA and RHMC, the fermion forces behave smoothly in all trajectories. 
However, the fermion forces of EOFA are substantially smaller than their counterparts in RHMC.
The averages of the maximum fermion forces are: 

\begin{center}
  \begin{tabular}{ccccccc}
\toprule
  & \multicolumn{4}{c}{EOFA} & \multicolumn{2}{c}{RHMC} \\ 
  & $(F_{\phi_1})_{\rm light}$ & $(F_{\phi_2})_{\rm light} $ 
  & $(F_{\phi_1})_{\rm heavy}$ & $(F_{\phi_2})_{\rm heavy} $ 
  & $(F)_{\rm light}$ & $(F)_{\rm heavy}$ \\
Conventional DWF & 0.0046(1) & 0.0331(2) & 0.0609(1) & 0.1318(2) & 0.1076(1) & 0.2855(1) \\
\midrule
     Optimal DWF & 0.0009(2) & 0.0139(2) & 0.0487(1) & 0.1810(1) & 0.0695(3) & 0.3534(1) \\ 
\bottomrule
  \end{tabular}
\end{center}

Note that for EOFA, the fermion forces of $\phi_1$ are much smaller than their counterparts of $\phi_2$. 
This immediately implies that $\phi_1$ and $\phi_2$ can be updated at two different time scales. 
This will be exploited in the tests on the $ 16^3 \times 32 \times 16 $ lattice. 

For tests of $N_f = 1 $ QCD on the $ 8^3 \times 16 \times 16 $ lattice, we set the multiple-time scales as follows.
With the length of the HMC trajectory equal to one, 
three different time scales are set to $\{k_0,k_1,k_2 \}=\{1,1,10\}$, and the fields are updated according to the following assignment: 
\BAN
&&k_0 : U_{\mu} (\textmd{gauge field}), \\
&&k_1 : \phi_1(\textmd{EOFA, heavy fermion}), \phi_2(\textmd{EOFA, heavy fermion}), \phi(\textmd{RHMC, heavy fermion}), \\
&&k_2 : \phi_1(\textmd{EOFA, light fermion}), \phi_2(\textmd{EOFA, light fermion}), \phi(\textmd{RHMC, light fermion}). \\
\EAN
Thus the smallest time interval in the molecular dynamic is $1/(k_0 k_1 k_2)$, and 
the numbers of momentum updates for $\{k_0,k_1,k_2\}$
are $\{8 k_0 k_1 k_2+1,4 k_1 k_2+1,2 k_2+1\}$ respectively, according to the Omelyan integrator.

Using one core of Intel i7-3820 CPU@3.60GHz,  
we measure the average time per HMC trajectory (T) and the acceptance rate (A) after thermalization, and obtain
the following results.

\begin{center}
  \begin{tabular}{ccccc}
\toprule
  & \multicolumn{2}{c}{EOFA} & \multicolumn{2}{c}{RHMC} \\ 
  & T (seconds) & A & T (seconds) & A \\
Conventional DWF &  6293(77) & 0.980(9) & 7365(96) & 0.996(4) \\ 
\midrule
     Optimal DWF &  8916(263) & 0.980(9) & 10657(538) & 0.984(8) \\
\bottomrule
  \end{tabular}
\end{center}

Thus, in both cases (conventional DWF and optimal DWF), EOFA outperforms RHMC 
for $N_f = 1 $ QCD on the $ 8^3 \times 16 \times 16 $ lattice.

\subsection{$L^3\times T = 16^3\times 32 $}

Next we turn to tests of $ N_f = 1 $ and $ N_f = (2+1) $ QCD on the $16^3\times 32 \times 16$ lattice, 
for the conventional DWF with kernel $ H = H_w (2 + \gamma_5 H_w)^{-1} $. 
The details of the simulation of 2-flavors of DWF have been presented in Ref. \cite{Chiu:2013aaa}.
After the initial thermalization of 300 trajectories (done with a GPU), 
we pick one configuration and use 4 cores CPU of i7-4820K CPU@3.70GHz 
to continue the HMC simulation with EOFA and RHMC respectively, and accumulate 5 trajectories in each case. 
With the length of the HMC trajectory equal to one, 
four different time scales are set to $\{k_0, k_1, k_2, k_3 \}=\{10, 1, 3, 2 \}$, and the fields are updated according to  
the following assignment:
\BAN
&&k_0 : U_{\mu} (\textmd{gauge field}), \\
&&k_1 : \phi_2(\textmd{EOFA, heavy fermion}), \phi(\textmd{RHMC, heavy fermion}), \\
&&k_2 : \phi_1(\textmd{EOFA, heavy fermion}), \phi_2(\textmd{EOFA, light fermion}), \phi(\textmd{RHMC, light fermion}), \\
&&k_3 : \phi_1(\textmd{EOFA, light fermion}).
\EAN
Then the smallest time interval in the molecular dynamic is $1/(k_0 k_1 k_2 k_3)$, 
and the numbers of momentum updates for $\{k_0,k_1,k_2,k_3\}$
are $\{16k_0k_1k_2k_3+1,8k_1k_2k_3+1,4k_2k_3+1,2k_3+1\}$ respectively, 
according to the Omelyan integrator.

With the statistics of five trajectories (all accepted), 
the average time (seconds) for generating one HMC trajectory (after thermalization) is listed below.

\begin{center}
\begin{tabular}{|c|cc|}
\hline
                 &    EOFA                &   RHMC      \\
\hline
$ N_f = 1 $      & 93241(290)             & 119445(408)         \\
$ N_f = 2+1 $    & 143099(833)            & 172569(588)         \\
\hline
\end{tabular}
\end{center}

These results suggest that EOFA outperforms RHMC for $N_f=1$ and $N_f=(2+1)$ QCD with the conventional DWF.

\section{Conclusion}

In this paper, we compare the performances of EOFA and RHMC, for $ N_f = 1 $ and $ N_f = 2 + 1 $ QCD
with DWF, on the $ 8^3 \times 16 \times 16 $ and $ 16^3 \times 32 \times 16 $ lattices respectively. 
Our results suggest that EOFA outperforms RHMC, no matter in terms of the computational efficiency or  
the memory requirement. This makes EOFA a better choice for dynamical simulations of lattice QCD with DWF.  
Currently, TWQCD Collaboration is using EOFA to simulate lattice QCD with $ (u,d,s,c) $ quarks
on the $ 24^3 \times 48 \times 16 $ and $ 32^3 \times 64 \times 16 $ lattices, with Nvidia GPUs (GTX-TITAN).

\begin{acknowledgments}
  This work is supported in part by the Ministry of Science and Technology
  (No.~NSC102-2112-M-002-019-MY3) and NTU-CQSE (Nos.~103R891404).
\end{acknowledgments}

\end{document}